# Hysteretic "Magnetic-Transport-Structural" Transition in "114" Cobaltites: Size Mismatch Effect

Tapati Sarkar*, V. Caignaert, V. Pralong and B. Raveau

*Laboratoire CRISMAT, UMR 6508 CNRS ENSICAEN,*
*6 bd Maréchal Juin, 14050 CAEN, France*

**Abstract**

The triple "magnetic-transport-structural" transition versus temperature in three series of "114" cobaltites – $Y_{1-x}Yb_xBaCo_4O_7$, $Y_{1-x}Ca_xBaCo_4O_7$ and $Yb_{1-x}Ca_xBaCo_4O_7$ – has been studied using magnetic, transport and differential scanning calorimetric measurements. The effect of the size mismatch $\sigma^2$, due to cationic disordering at the Ln sites upon such a transition is shown for the first time in a triangular lattice. We show that increasing $<r_{Ln}>$ has an effect of increasing $T_S$ dramatically, while the size mismatch $\sigma^2$ at the Ln sites decreases $T_S$ substantially. Moreover, the cationic mismatch at the Ln sites modifies the nature of the hysteretic transition by turning the sharp first order transition seen in the undoped samples into an intermix of first and second order transitions. These results are discussed on the basis of the particular nature of the high temperature form which exhibits a hexagonal close packed structure (space group: $P6_3mc$) with respect to the low temperature orthorhombic form (space group: $Pbn2_1$), the latter corresponding to a distortion of the former due to a puckering of the kagomé layers.

* Corresponding author: Dr. Tapati Sarkar
e-mail: tapati.sarkar@ensicaen.fr
Fax: +33 2 31 95 16 00
Tel: +33 2 31 45 26 32

**Introduction**

The recent studies of the "114" LnBaCo$_4$O$_7$ cobaltites [1 – 6], which are isostructural to the oxide LnBaZn$_3$AlO$_7$, previously discovered by Mueller-Buschbaum et. al. [7], have generated a lot of interest in the scientific community due to their complex magnetic, electronic, thermoelectric and electrochemical properties. In these oxides, the barium and oxygen atoms form a close packed stacking, whereas the cobalt and lanthanide elements occupy the tetrahedral and octahedral sites of this framework respectively. These oxides are one of the very rare examples of pure tetrahedral frameworks, where the transition element, in this case, cobalt, exhibits a mixed valence i.e. Co$^{2+}$/Co$^{3+}$, leading to the description of the 3D "Co$_4$O$_7$" framework as the stacking of two sorts of layers of corner shared CoO$_4$ tetrahedra, the triangular (T) layers and kagomé (K) layers (**Fig. 1**).

One intriguing and original feature of these compounds deals with the fact that they exhibit a structural transition at a characteristic temperature ($T_S$) which also coincides with a hysteretic magnetic and electrical transport transition. Indeed, at room temperature or slightly above room temperature these oxides exist in the hexagonal *P6$_3$mc* (or trigonal *P31c*) symmetry (**Fig. 1(a)**), and by decreasing the temperature, an orthorhombic *Pbn2$_1$* symmetry is obtained below $T_S$ [8 – 12]. This structural transition corresponds to a distortion of the room temperature framework (**Fig. 1(b)**). Simultaneously, these oxides show a jump of resistivity and a drop in the magnetization around $T_S$ as the temperature decreases, and moreover this phenomenon is hysteretic. The origin of this triple transition is still a matter of debate. It has been attributed by several authors to a possible redistribution of the charges of cobalt cations at $T_S$ [8, 13], whereas other reports have indicated that the transition occurs as a response to a severely underbonded Ba$^{2+}$ cation in the high temperature hexagonal phase [9] rather than any charge ordering of Co$^{2+}$/Co$^{3+}$ species. However, the recent study on the "114" cobaltite CaBaCo$_4$O$_7$ [14], which shows that in spite of its large orthorhombic distortion this phase still exhibits a strongly underbonded Ba$^{2+}$ cation does not support the latter statement.

At this point of the investigations, it is most important to consider that the structural transition may be the driving force which is at the origin of the magnetic and electron transport transition, similar to what happens in colossal magnetoresistive (CMR) manganites [15]. Indeed, it has been observed by several authors that the transition temperature $T_S$ increases with the size of the Ln$^{3+}$ cation [10, 12, Fig. 2 or Ref. 16] but no real explanation was given about this phenomenon. Considering the close packed character of the hexagonal high temperature form of these compounds, which involves a *c/a* ratio close to (but slightly different from) the ideal one, i.e. $\frac{c}{a} = \sqrt{\frac{8}{3}}$, we believe that these compounds lie at the borderline of a puckering of the "$O_4$" and "$BaO_3$" layers, which can then be induced by decreasing the temperature, thereby leading to the orthorhombic distortion. This would explain why $T_S$ is very sensitive to the size of the $Ln^{3+}$ cations located in the octahedral cavities of the close packed framework "$BaO_7$". The phase transition, being first order in nature, leaves its signature on the physical properties of the system, and can be detected by temperature dependent measurements of magnetization and electrical resistivity, involving the development of a different magnetic component and concomitantly charge localization just below $T_S$, as previously shown for $YbBaCo_4O_{7+\delta}$ [17].

In the above scenario, the size of the $Ln^{3+}$ cations, as well as their size difference (when the octahedral sites are occupied by different $Ln^{3+}$ cations simultaneously) should dramatically influence this triple structural-magnetic-electrical transport transition, as previously observed for manganites [15, 18]. Thus, in this paper we have investigated this triple transition for three series of oxides, $Y_{1-x}Yb_xBaCo_4O_7$, $Y_{1-x}Ca_xBaCo_4O_7$ and $Yb_{1-x}Ca_xBaCo_4O_7$, and we have shown that in addition to the size effect of the $Ln^{3+}$ cations, their size mismatch, previously defined as $\sigma^2$ [18], plays a key role in determining the transition temperature as well as the nature of the hysteretic transition.

**Experimental**

Phase-pure samples of $Y_{1-x}Yb_xBaCo_4O_7$ [x = 0 − 1] and $(Y/Yb)_{1-x}Ca_xBaCo_4O_7$ [x = 0 − 0.3] were prepared by solid state reaction technique. The precursors used were $Y_2O_3$, $Yb_2O_3$, $CaCO_3$, $BaCO_3$ and $Co_3O_4$. In a first step, a stoichiometric mixture of the required precursors was intimately ground and heated at 900°C in air for 12 hrs for decarbonation. The mixture was then pressed in the form of rectangular bars, annealed at 1150°C for 12 hrs and finally quenched down to room temperature. The phase purity of the samples was checked using X-ray diffraction (XRD). No impurity phases were detected. The X-ray diffraction patterns were registered with a Panalytical X'Pert Pro diffractometer with a Cu-Kα source (λ = 1.54 Å) under a continuous scanning mode in the 2θ range 10° - 120° and step size Δ2θ = 0.017°.

The as-prepared samples had a slight oxygen excess [$A_{1-x}A'_xBaCo_4O_{7+\delta}$, with δ ≈ 0.05]. As such, in a final step, the as-prepared samples were annealed at 600°C for 2 hrs in an argon atmosphere in order to obtain oxygen stoichiometric samples. The oxygen content of the as-prepared as well as the argon annealed samples was determined by iodometric titration.

Around 20 mg of the sample to be studied was dissolved under argon flow in 50 ml of acetic buffer solution containing an excess of KI (~ 200 mg), resulting in a reduction of the tri- and tetra-valent cobalt species of the sample to $Co^{2+}$ ions, and formation of a stoichiometric amount of iodine in the solution. Iodine was then titrated with 0.1 M $Na_2S_2O_3$ solution using thiodene as the indicator. The end-point was detected visually as the yellow coloured solution turned colourless.

The d.c. magnetization measurements were performed using a superconducting quantum interference device (SQUID) magnetometer with variable temperature cryostat (Quantum Design, San Diego, USA). Electrical resistivity was measured from 5 K to 325 K using the four-probe d.c. technique in a physical property measurement system (PPMS) from Quantum Design. All the magnetic and electrical transport properties were registered on dense ceramic bars of dimensions ~ $4 \times 2 \times 2$ $mm^3$. The differential scanning calorimetric (DSC) measurements were performed at a rate of 10 K/min with a SDT 2920 TA Instruments.

**Results and discussion**

*Structural Study*

Considering the structural results obtained by previous authors [10 – 12, Fig. 2 of Ref. 16, 19] vis-à-vis the LnBaCo$_4$O$_7$ series, it can be clearly observed from the approximately linear dependence of the structural transition temperature $T_S$ versus the effective ionic radius of $Ln^{3+}$ [20] (**Fig. 2(a)**) that the stability of the orthorhombic structure increases dramatically at the expense of the hexagonal one as the size of the $Ln^{3+}$ cation increases. [Note: The presence of a structural transition in HoBaCo$_4$O$_7$ is still a matter of debate, with Ref. 11 reporting a clear structural transition at T = 355 K, while Ref. 16 states that HoBaCo$_4$O$_7$ does not show any phase transition]. The fact that the hexagonal phase (**Fig. 1(a)**) exhibits a close packing of the BaO$_3$ and O$_4$ layers, and that the orthorhombic phase (**Fig. 1(b)**) is a distortion of the hexagonal structure involving a puckering of the kagomé layers, suggests that the structural transition temperature $T_S$ is very sensitive to the deviation from the ideal hexagonal close packing. Under these conditions, the size of the $Ln^{3+}$ cations, which are located in the octahedral positions, will play a crucial role in this transition i.e. beyond a critical size of the $Ln^{3+}$ cation, the close packed stacking may be destroyed, thereby leading to the orthorhombic symmetry. In order to understand this phenomenon, we have plotted the *c/a* values of the room temperature (RT) hexagonal phases (or the equivalent *c/a'* values of the RT orthorhombic phases, where, $a' = \left(\dfrac{a + b/\sqrt{3}}{2}\right)$) versus $\langle r_{Ln}\rangle$ (**Fig. 2(b)**). It can be seen from **Fig. 2(b)** that the $c/a$ values of the RT hexagonal oxides (Lu – Er) are very close to the ideal value (1.633) of the close packed structure, decreasing almost linearly from 1.633 to 1.631 as the size of the $Ln^{3+}$ cation increases from 0.861 Å (for $Lu^{3+}$) to 0.890 Å (for $Er^{3+}$). On increasing the cation radius further, a sudden collapse of the equivalent $c/a'$ values of the RT orthorhombic oxides is observed for $r_{Ln} > 0.890$ Å down to 1.624 for $Y^{3+}$ ($r_Y$ = 0.900 Å) and 1.620 for $Ho^{3+}$ ($r_{Ho}$ = 0.901 Å). This demonstrates that there exists a critical value of the ionic radius of $Ln^{3+}$ ($r_C$) ~ 0.890 Å beyond which the hexagonal close packed structure is destroyed at the benefit of the orthorhombic phase. It also suggests that the hexagonal – orthorhombic phase transition is first order in nature. The comparison of **Fig. 2(a) and Fig. 2(b)** further indicates that a small deviation of the structure from the ideal $c/a$ close packing tends to destabilize the hexagonal form at the benefit of the orthorhombic structure. This can be seen for the Lu to Er series, where $T_S$ increases from 160 K for Lu (which exhibits an almost perfect hexagonal close packing at room temperature) to 288 K for Er. From this analysis, it appears that all the samples with $r_{Ln} < 0.890$ Å exhibit a $c/a$ value larger than 1.631, and are hence predicted to be hexagonal at room temperature, whereas the orthorhombic distortion should be observed for $r_{Ln} > 0.900$ Å for which the $c/a'$ values at room temperature are smaller than 1.625. The zone of $r_{Ln}$ values lying between 0.890 Å and 0.900 Å (the square shaded region in **Fig. 2(b)**) remains an open issue.

**Tables 1**, **2** and **3** give a comprehensive list of all the samples that have been explored in the present study. The room temperature symmetry and cell parameters obtained from the Rietveld analysis of the X-ray diffraction patterns using the FULLPROF refinement program [21] have been listed, together with the average ionic radii of the $Ln^{3+}$ cations according to Shannon [20] and computed for the sample having the generic formula $A_{1-x}A'_xBaCo_4O_7$ using the simple formula $\langle r_{Ln}\rangle = (1-x)*\langle r_A\rangle + x*\langle r_{A'}\rangle$. The reliability factors, also listed in the Tables, confirm the validity of the structural data.

The results are in accordance with the above predictions for most of the synthesized compounds. From **Tables 1**, **2** and **3**, and from the $c/a$ and $c/a'$ values plotted for these oxides in **Fig. 2(b)**, we observe that all the compounds of the series $Y_{1-x}Yb_xBaCo_4O_7$ which have $r_{Ln} < r_C$ are hexagonal at room temperature with a $c/a$ value larger than or close to 1.630 (black open squares in **Fig. 2(b)**). On the other hand, the $Y_{1-x}Ca_xBaCo_4O_7$ oxides whose $\langle r_{Ln}\rangle$ values are larger than $r_C$ exhibit the orthorhombic symmetry with a $c/a'$ value smaller than 1.625 (red open circles in **Fig. 2(b)**). However, the case of the $Yb_{1-x}Ca_xBaCo_4O_7$ series is more complex. Here, for small substitution degree (x ≤ 0.10), the hexagonal structure with $c/a \approx 1.630$ is stabilized in accordance with the fact that $<r_{Ln}> = 0.881$ Å which is smaller than $r_C$. For a larger substitution degree (x = 0.20), the phase is still hexagonal with a $c/a$ value 1.629 corresponding to a $<r_{Ln}>$ value of 0.894 Å which is located in the "open issue zone", suggesting that the hexagonal close packing may be stabilized for $c/a$ value down to 1.629. Finally, for x = 0.30, we observe a peculiar behaviour, since the symmetry is hexagonal but with a $c/a$ value of 1.627 (marked by a shaded circle in **Fig. 2(b)**), whereas, according to our predictions, it should have been orthorhombic as per its $<r_{Ln}>$ value of 0.908 Å. This exceptional behaviour of $Yb_{0.7}Ca_{0.3}BaCo_4O_7$ might be due to the fact that the mixed valence of cobalt may play a role in the stabilization of the hexagonal phase. This would explain the different behaviour of this sample compared to that of the $Y_{1-x}Yb_xBaCo_4O_7$ series, but not with respect to that of the $Y_{1-x}Ca_xBaCo_4O_7$ series, since both the oxides, $Y_{0.7}Ca_{0.3}BaCo_4O_7$ and $Yb_{0.7}Ca_{0.3}BaCo_4O_7$ exhibit the same cobalt valence. It also suggests that the size difference between $Yb^{3+}$ and $Ca^{2+}$ cations which is much larger than that between $Y^{3+}$ and $Ca^{2+}$ cations plays an important role in the stabilization of the hexagonal phase with respect to the orthorhombic phase for $Yb_{0.7}Ca_{0.3}BaCo_4O_7$.

In order to evaluate the role of the A-site size mismatch in this structural transition, we have calculated the variance $\sigma^2$ of the different samples. The variance $\sigma^2$ quantifies the random disorder of $A^{3+}$ and $B^{3+/2+}$ cations with different sizes distributed over the $A$-sites. Thus, for two $A$-site species with fractional occupancies $x_i$ $\left(\sum x_i = 1\right)$, the variance of the ionic radii about the mean radius ($<r_{Ln}>$) has been defined as $\sigma^2 = \sum x_i r_i^2 - \langle r_{Ln} \rangle^2$. $\sigma^2$, also listed in **Tables 1**, **2** and **3**, is very important in the context of the present study. We note that the highest value of $\sigma^2$ is indeed observed for $Yb_{0.7}Ca_{0.3}BaCo_4O_7$, suggesting that the size mismatch stabilizes the hexagonal close packing at the expense of the orthorhombic structure. The influence of the size mismatch upon the evolution of the structural transition temperature $T_S$ of these oxides will be studied further in the subsequent sections.

The structural phase transition that we are trying to probe in this paper leaves its signature on the physical properties (magnetization as well as electrical transport) of the system. It can also be seen in differential scanning calorimetric (DSC) measurements as an endothermic (or exothermic) peak in the heating (or cooling) curves. As such, in the following sections, we probe the d.c. magnetization, electrical resistivity and the DSC curves for the samples.

*D. C. magnetization study*

With the specific aim of probing the nature of the structural phase transition, the temperature dependence of d.c. magnetization was measured following two protocols, namely field cooled cooling (FCC) and field cooled warming (FCW). In the FCC mode, an applied magnetic field of 0.3 T was switched on at T = 400 K (which lies above the transition temperature for all the measured samples), and the measurement was made while cooling the sample down to T = 5 K. On reaching 5 K in the FCC mode, the data were taken again in the presence of the same magnetic field while warming up the sample. This was the FCW mode.

As an example, we show the M-T data for $YBaCo_4O_7$ in **Fig. 3**. The thermal hysteresis seen in the FCC and FCW data (marked by a red circle and also enlarged and shown in the inset of **Fig. 3**) occurs as a magnetic response to the structural phase transition occurring in $YBaCo_4O_7$ at ~ 300 K[10]. Besides the hysteretic character of the transition, note its sharpness, which illustrates its strongly first order nature. Since in this paper we deal exclusively with the nature of the structural phase transition, for the remaining samples, the magnetic data will be shown only around the temperature range in which the structural phase transition occurs, instead of the full measured temperature range of 5 K – 400 K.

**Fig. 4 (a)** shows the M-T curves for the samples listed in **Table 1**. As can be seen from **Table 1**, the average cation radius ($<r_{Ln}>$) increases from 0.868 Å to 0.900 Å as we move down the five panels of **Fig. 4 (a)**. Correspondingly, the structural transition temperature ($T_S$) also shows a systematic increase with increase in $<r_{Ln}>$. This is in accordance with earlier reports of the dependence of $T_S$ on $r_{Ln}$[8, 12, Fig. 2 of Ref. 16]. However, an additional feature that becomes immediately evident from **Fig. 4 (a)** is that for the samples with intermediate compositions between $YbBaCo_4O_7$ and $YBaCo_4O_7$ containing two different cations, $Yb^{3+}$ and $Y^{3+}$, in the same octahedral site, the transition is much less sharp making it more difficult to determine $T_S$ accurately. Such a flattened and broader transition can be explained by the existence of structural disordering between $Yb^{3+}$ and $Y^{3+}$ cations in the octahedral sites of the structure. A similar smearing and rounding-off of a first order magnetic phase transition has been observed in manganites [22] due to cationic disordering i.e. size mismatch between the A-site cations. Thus, these results show that size mismatch of $Ln^{3+}$ cations modifies the nature of the phase transition dramatically i.e. samples having no mismatch with $\sigma^2 = 0$ (top and bottom panel of **Fig. 4(a)**) show a sharp transition, whereas samples involving size mismatch with $\sigma^2 \neq 0$ exhibit a smeared and rounded off transition (middle three panels of **Fig. 4(a)**).

A similar evolution of the M(T) curves versus composition is observed for the oxides $Y_{1-x}Ca_xBaCo_4O_7$ (**Fig. 4(b)**) and $Yb_{1-x}Ca_xBaCo_4O_7$ (**Fig. 4(c)**). In both these cases, the transition temperature is also characterized by a hysteresis of the FCC and FCW curves and shifts to a higher value as x increases, in accordance with the larger cation radius of $Ca^{2+}$ vis-à-vis that of $Yb^{3+}$ and $Y^{3+}$. However, by substituting $Ca^{2+}$ at the $Y^{3+}$ or $Yb^{3+}$ site, we increase the degree of disorder further. This can be seen from the $\sigma^2$ values for these compounds (listed in **Tables 2** and **3**) which are one order of magnitude larger than the $\sigma^2$ values for the series $Y_{1-x}Yb_xBaCo_4O_7$ (**Table 1**). As a result of this higher size mismatch a complete rounding-off of the first order transition is rapidly reached. Thus, for x > 0.2, no thermal hysteresis is observed in the FCC and FCW curves implying a complete suppression of the first order phase transition in the sample.

*Electrical transport study*

Electrical transport under an applied field of 0.3 T was measured using the same protocol (FCC and FCW) as that used in the M-T measurements, except that the measured physical quantity was resistance in this case. The data obtained for the samples listed in **Tables 1**, **2** and **3** are shown in **Fig. 5 (a)**, **(b)** and **(c)** respectively.

The results obtained from the ρ-T measurements are qualitatively similar and support our conclusions as derived from the M-T measurements in the previous section. The resistive transition temperature is close to that of the magnetic transition and increases with an increase in the average cationic radius $<r_{Ln}>$. Moreover, the height of the jump of resistivity at $T_S$ decreases and the hysteretic transition is broadened as cationic disorder is introduced i.e. as $\sigma^2$ increases. Note that the samples with x > 0.2 in the $(Y/Yb)_{1-x}Ca_xBaCo_4O_7$ series show an absence of thermal hysteresis in the $\rho_{FCC}(T)$ and $\rho_{FCW}(T)$ data also (not shown here) similar to their $M_{FCC}(T)$ and $M_{FCW}(T)$ behaviour.

*Differential Scanning Calorimetry (DSC) Study: Influence of Size mismatch upon $T_S$*

Magnetization and resistivity measurements clearly show the effect of cationic disordering or size mismatch $\sigma^2$ of $Ln^{3+}/Ca^{2+}$ cations upon the nature of the structural transition, showing a broadening and rounding-off of the hysteretic curves at the transition with increasing disorder. However, they do not allow the transition temperature to be determined with accuracy because of the very broad character of the curves as $\sigma^2$ increases. In this respect, DSC measurements should allow $T_S$ to be determined with a much higher accuracy. We have thus recorded DSC curves for the three series of samples. The data were recorded while warming the samples. For each of these samples the phase transition temperatures are clearly visible as endothermic peaks (**Fig. 6**), and the transition temperatures correspond well with the values of $T_S$ deduced from magnetization and resistivity measurements. More importantly, it is worth pointing out that the non-substituted samples, $YBaCo_4O_7$ and $YbBaCo_4O_7$, which have no mismatch ($\sigma^2 = 0$), exhibit well resolved peaks corresponding to large latent heat, whereas for the A-site substituted samples, involving non-zero values of $\sigma^2$, the peaks are much smaller and broader, indicating that the latent heat associated with the transition is reduced by increasing the size mismatch $\sigma^2$, till it gets completely suppressed in the samples with the highest degree of disorder. Note that for the $Y_{1-x}Ca_xBaCo_4O_7$ samples (**Fig. 6(b)**) and $Yb_{1-x}Ca_xBaCo_4O_7$ ((**Fig. 6(c)**), the DSC curves tend to flatten rapidly as x reaches 0.20, due to the higher mismatch values of $\sigma^2$. For x > 0.20, the DSC curves (not shown here) show no change of slope that would signify a transition.

The important influence of the size mismatch upon the structural transition temperature of these oxides is illustrated by comparing the DSC curves (**Fig. 7(a)**) as well as the resistivity curves (**Fig. 7(b)** and **Fig. 7(c)**) of two samples having practically the same ionic radius $<r_{Ln}>$, $Y_{0.7}Yb_{0.3}BaCo_4O_7$ and $ErBaCo_4O_7$ ($<r_{Ln}> = 0.890$ Å), but the former exhibiting a cationic disorder ($\sigma^2 = 2.15 \times 10^{-4}$ Å$^2$) and the latter having no disorder ($\sigma^2 = 0$).

As is evident from **Fig. 7**, the $T_S$ value of $Y_{0.7}Yb_{0.3}BaCo_4O_7$ (258.3 K), for which $\sigma^2 = 2.15 \times 10^{-4}$ Å$^2$, is smaller by 30 K than that of $ErBaCo_4O_7$ (288.2 K), for which $\sigma^2 = 0$, both oxides having the same value of $<r_{Ln}>$. This demonstrates that the size mismatch in the lanthanide sites causes the structural transition temperature to decrease dramatically thereby stabilizing the hexagonal phase at the expense of the orthorhombic form. This effect of the size mismatch on $T_S$ can be generalized to all the oxides of the three series as shown from their $T_S$ values plotted in **Fig. 2 (a)**. We can see that the $T_S$ values of all the samples of the three series containing two different cations on the Ln site (open symbols) are much below the dashed line corresponding to the $T_S$ values of the $LnBaCo_4O_7$ series. Thus, for a given value of $<r_{Ln}>$, the $T_S$ value of the $Ln_{1-x}A_xBaCo_4O_7$ cobaltites are much smaller than expected from the line given for the $LnBaCo_4O_7$ oxides. This effect is especially high for samples with higher values of $\sigma^2$, as shown for instance for $Yb_{0.8}Ca_{0.2}BaCo_4O_7$, for which a $T_S$ value of ~ 300 K should be expected in the absence of size mismatch, whereas the actual value of $T_S$ for this particular sample is only 207 K due to its high $\sigma^2$ value.

In fact we find that the reduction in $T_S$ for the disordered samples ($\Delta T_S$) is proportional to the square root of $\sigma^2$ i.e. $\Delta T_S \propto \sqrt{\sigma^2}$. We calculate the proportionality constant from the value $\Delta T_S$ for the samples $Y_{0.7}Yb_{0.3}BaCo_4O_7$ and $ErBaCo_4O_7$ which have the same $<r_{Ln}>$, and then estimate the $\Delta T_S$ values for all the other disordered samples. This is also shown in **Fig. 2 (a)**. The upward arrows starting from the open symbols in **Fig. 2 (a)** and ending in the corresponding closed symbols indicate the corrected values of $T_S$ i.e. what should have been the values of $T_S$ for these samples if there was no disorder involved. As is quite clear from the figure, the corrected values of $T_S$ lie very close to the dashed line corresponding to the $T_S$ values of the $LnBaCo_4O_7$ series thereby confirming that the reduction in $T_S$ can indeed be attributed to the presence of disorder in these samples. This also gives us a quantitative way of estimating the transition temperature in any sample having cationic disorder.

**Conclusion**

In the present study, the effects of the size $<r_{Ln}>$ and of the size mismatch ($\sigma^2$) of the Ln site cations upon the triple magnetic-transport-structural transition of the "114" cobaltites $Ln_{1-x}A_xBaCo_4O_7$ have been studied and decoupled using three series of oxides as prototypical examples. We have shown that while the increase of the mean size $<r_{Ln}>$ of the $Ln^{3+}$ cations increases $T_S$ thereby stabilizing the orthorhombic form at the expense of the hexagonal form, the cationic size mismatch ($\sigma^2$) at the Ln site decreases $T_S$ thereby stabilizing the hexagonal form at the expense of the orthorhombic form, the magnetic and electrical transport transitions following the same pattern. Importantly, while variation in $<r_{Ln}>$ can vary only $T_S$, an increase in $\sigma^2$ affects the nature of the hysteretic transition on the M(T) and ρ(T) curves by smearing out its strongly first order nature seen in the ordered $LnBaCo_4O_7$ phases ($\sigma^2 = 0$). This behaviour of the "114" cobaltites exhibits a great similarity with that of other strongly correlated electron systems, such as the CMR manganites, whose magnetic-transport-structural transitions are also determined by size and size mismatch of A-site cations [15]. The combination of the two parameters, $<r_{Ln}>$ and $\sigma^2$, will be the key factor in the future for tuning the magnetic, transport and structural transitions of these "114" cobaltites.

**Acknowledgements**

The authors acknowledge the CNRS and the Conseil Regional of Basse Normandie for financial support in the frame of Emergence Program. V. P. acknowledges support by the ANR-09-JCJC-0017-01 (Ref: JC09_442369).

**Table captions**

**Table 1**: Unit cell parameters at room temperature, average cation radius ($<r_{Ln}>$), $c/a$ at room temperature and cation size variance ($\sigma^2$) for the series $Y_{1-x}Yb_xBaCo_4O_7$.

**Table 2**: Unit cell parameters at room temperature, average cation radius ($<r_{Ln}>$), $c/a$ at room temperature and cation size variance ($\sigma^2$) for the series $Y_{1-x}Ca_xBaCo_4O_7$.

**Table 3**: Unit cell parameters at room temperature, average cation radius ($<r_{Ln}>$), $c/a$ at room temperature and cation size variance ($\sigma^2$) for the series $Yb_{1-x}Ca_xBaCo_4O_7$.

**Figure captions**

**Figure 1:** Perspective view of the $LnBaCo_4O_7$ structure in the (a) hexagonal and (b) orthorhombic symmetry.

**Figure 2:** (a) $T_S$ vs $<r_{Ln}>$ and (b) $c/a$ (or $c/a'$) vs $<r_{Ln}>$ at room temperature for some members of the cobaltite "114" oxide family. The (H) and (O) beside each data point in (b) indicate whether the particular sample is hexagonal or orthorhombic at room temperature. The values of $T_S$ and $c/a$ for $LuBaCo_4O_7$ and $TmBaCo_4O_7$ have been quoted from Ref 8, and the values of $T_S$ and $c/a$ for $HoBaCo_4O_7$ have been quoted from Ref. 11.

**Figure 3:** $M_{FCC}(T)$ and $M_{FCW}(T)$ curves of $YBaCo_4O_7$ measured at H = 0.3 T. The empty symbols are for $M_{FCC}(T)$ and the solid symbols are for $M_{FCW}(T)$. The inset shows the $M_{FCC}(T)$ and $M_{FCW}(T)$ curves in the temperature range (280 K – 315 K) in which the structural phase transition occurs for this sample.

**Figure 4:** $M_{FCC}(T)$ and $M_{FCW}(T)$ curves of (a) $Y_{1-x}Yb_xBaCo_4O_7$, (b) $Y_{1-x}Ca_xBaCo_4O_7$ and (c) $Yb_{1-x}Ca_xBaCo_4O_7$ measured at H = 0.3 T. The empty symbols are for $M_{FCC}(T)$ and the solid symbols are for $M_{FCW}(T)$.

**Figure 5:** $\rho_{FCC}(T)$ and $\rho_{FCW}(T)$ curves of (a) $Y_{1-x}Yb_xBaCo_4O_7$, (b) $Y_{1-x}Ca_xBaCo_4O_7$ and (c) $Yb_{1-x}Ca_xBaCo_4O_7$ measured at H = 0.3 T. The empty symbols are for $\rho_{FCC}(T)$ and the solid symbols are for $\rho_{FCW}(T)$.

**Figure 6:** DSC curves of (a) $Y_{1-x}Yb_xBaCo_4O_7$, (b) $Y_{1-x}Ca_xBaCo_4O_7$ and (c) $Yb_{1-x}Ca_xBaCo_4O_7$ measured while warming the samples. The phase transition temperatures are marked by arrows.

**Figure 7:** (a) DSC curves for $Y_{0.7}Yb_{0.3}BaCo_4O_7$ and $ErBaCo_4O_7$ measured while warming the samples. The phase transition temperatures ($T_S$) are marked by arrows, and $\rho_{FCC}(T)$ and $\rho_{FCW}(T)$ curves of (b) $Y_{0.7}Yb_{0.3}BaCo_4O_7$ and (c) $ErBaCo_4O_7$ measured at H = 0.3 T. The empty symbols are for $\rho_{FCC}(T)$ and the solid symbols are for $\rho_{FCW}(T)$.

**Table 1**

| Compound | Crystal system (Space group) | Unit cell parameters | | Reliability factor ($R_F$) | $<r_{Ln}>$ (Å) | $c/a$ | $\sigma^2$ (Å$^2$) |
|---|---|---|---|---|---|---|---|
| | | $a$ (Å) | $c$ (Å) | | | | |
| YBaCo$_4$O$_7$ | Orthorhombic ($Pbn2_1$) | 6.297(1) | $b$ = 10.919(1) $c$ = 10.230(1) | 6.96 | 0.900 | 1.624 | 0 |
| Y$_{0.7}$Yb$_{0.3}$BaCo$_4$O$_7$ | Hexagonal ($P6_3mc$) | 6.289(1) | 10.251(1) | 5.25 | 0.890 | 1.630 | 2.15 × 10$^{-4}$ |
| Y$_{0.5}$Yb$_{0.5}$BaCo$_4$O$_7$ | Hexagonal ($P6_3mc$) | 6.282(1) | 10.245(1) | 5.06 | 0.884 | 1.631 | 2.56 × 10$^{-4}$ |
| Y$_{0.3}$Yb$_{0.7}$BaCo$_4$O$_7$ | Hexagonal ($P6_3mc$) | 6.276(1) | 10.241(1) | 4.68 | 0.882 | 1.632 | 2.15 × 10$^{-4}$ |
| YbBaCo$_4$O$_7$ | Hexagonal ($P6_3mc$) | 6.269(1) | 10.231(1) | 5.80 | 0.868 | 1.632 | 0 |

**Table 2**

| Compound | Crystal system (Space group) | Unit cell parameters | | Reliability factor ($R_F$) | $\langle r_{Ln}\rangle$ (Å) | $c/a$ | $\sigma^2$ (Å$^2$) |
|---|---|---|---|---|---|---|---|
| | | $a$ (Å) | $c$ (Å) | | | | |
| YBaCo$_4$O$_7$ | Orthorhombic ($Pbn2_1$) | 6.297(1) | $b = 10.919(1)$ $c = 10.230(1)$ | 6.96 | 0.900 | 1.624 | 0 |
| Y$_{0.9}$Ca$_{0.1}$BaCo$_4$O$_7$ | Orthorhombic ($Pbn2_1$) | 6.298(1) | $b = 10.917(1)$ $c = 10.233(1)$ | 7.92 | 0.910 | 1.624 | $9.00 \times 10^{-4}$ |
| Y$_{0.8}$Ca$_{0.2}$BaCo$_4$O$_7$ | Orthorhombic ($Pbn2_1$) | 6.297(1) | $b = 10.935(1)$ $c = 10.228(1)$ | 10.5 | 0.920 | 1.622 | $1.60 \times 10^{-3}$ |
| Y$_{0.7}$Ca$_{0.3}$BaCo$_4$O$_7$ | Orthorhombic ($Pbn2_1$) | 6.297(1) | $b = 10.934(1)$ $c = 10.229(1)$ | 9.15 | 0.930 | 1.622 | $2.10 \times 10^{-3}$ |

**Table 3**

| Compound | Crystal system (Space group) | Unit cell parameters | | Reliability factor ($R_F$) | $\langle r_{Ln} \rangle$ (Å) | $c/a$ | $\sigma^2$ (Å$^2$) |
|---|---|---|---|---|---|---|---|
| | | $a$ (Å) | $c$ (Å) | | | | |
| YbBaCo$_4$O$_7$ | Hexagonal ($P6_3mc$) | 6.269(1) | 10.231(1) | 5.80 | 0.868 | 1.632 | 0 |
| Yb$_{0.9}$Ca$_{0.1}$BaCo$_4$O$_7$ | Hexagonal ($P6_3mc$) | 6.274(1) | 10.229(1) | 5.59 | 0.881 | 1.630 | 1.57 × 10$^{-3}$ |
| Yb$_{0.8}$Ca$_{0.2}$BaCo$_4$O$_7$ | Hexagonal ($P6_3mc$) | 6.279(1) | 10.227(1) | 5.71 | 0.894 | 1.629 | 2.79 × 10$^{-3}$ |
| Yb$_{0.7}$Ca$_{0.3}$BaCo$_4$O$_7$ | Hexagonal ($P6_3mc$) | 6.284(1) | 10.225(1) | 5.85 | 0.908 | 1.627 | 3.66 × 10$^{-3}$ |

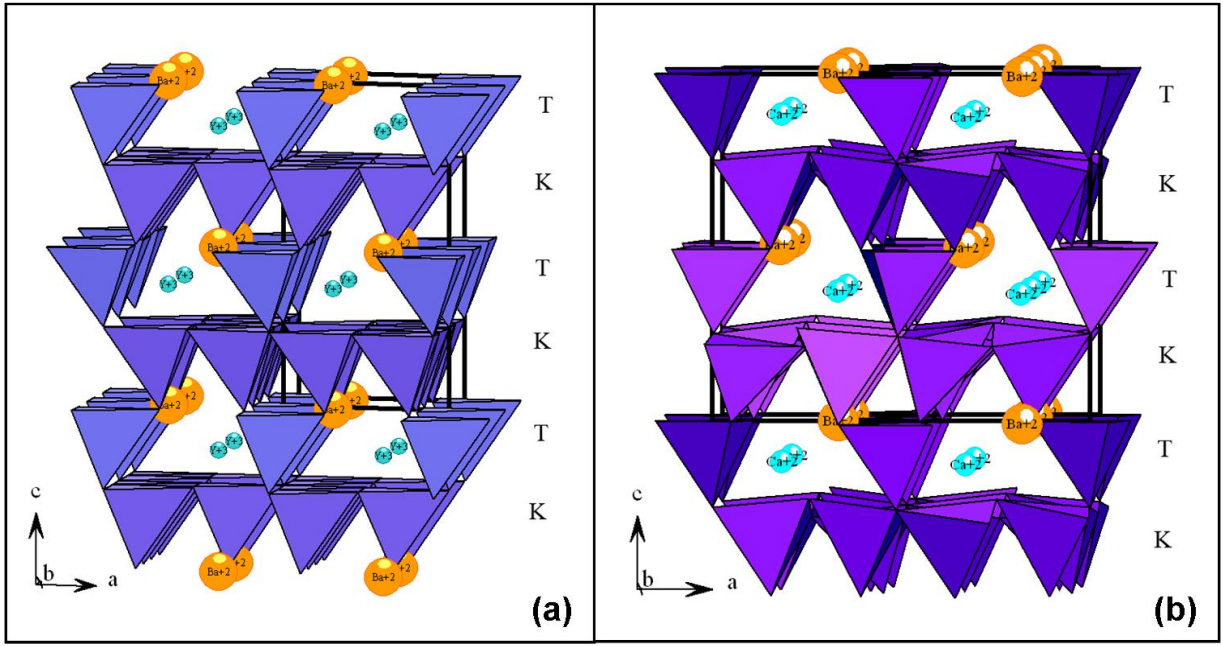

Figure 1

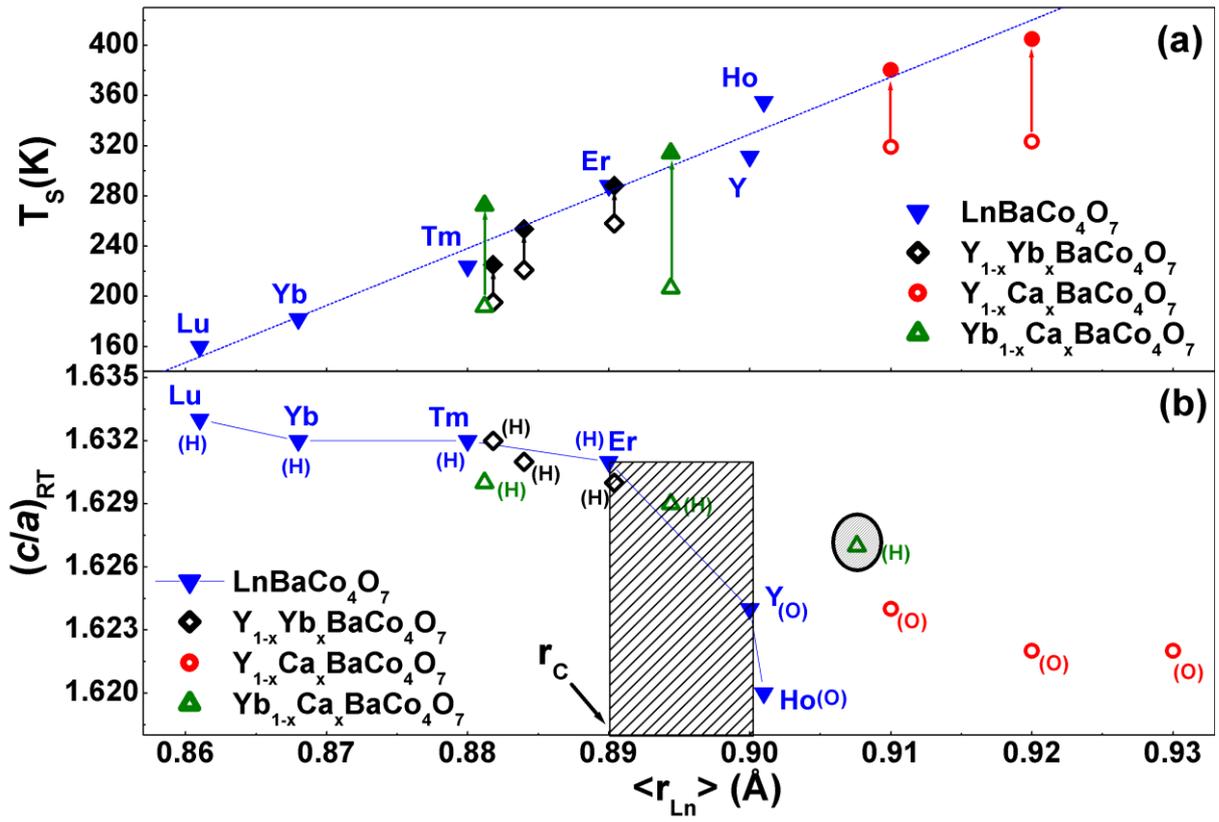

Figure 2

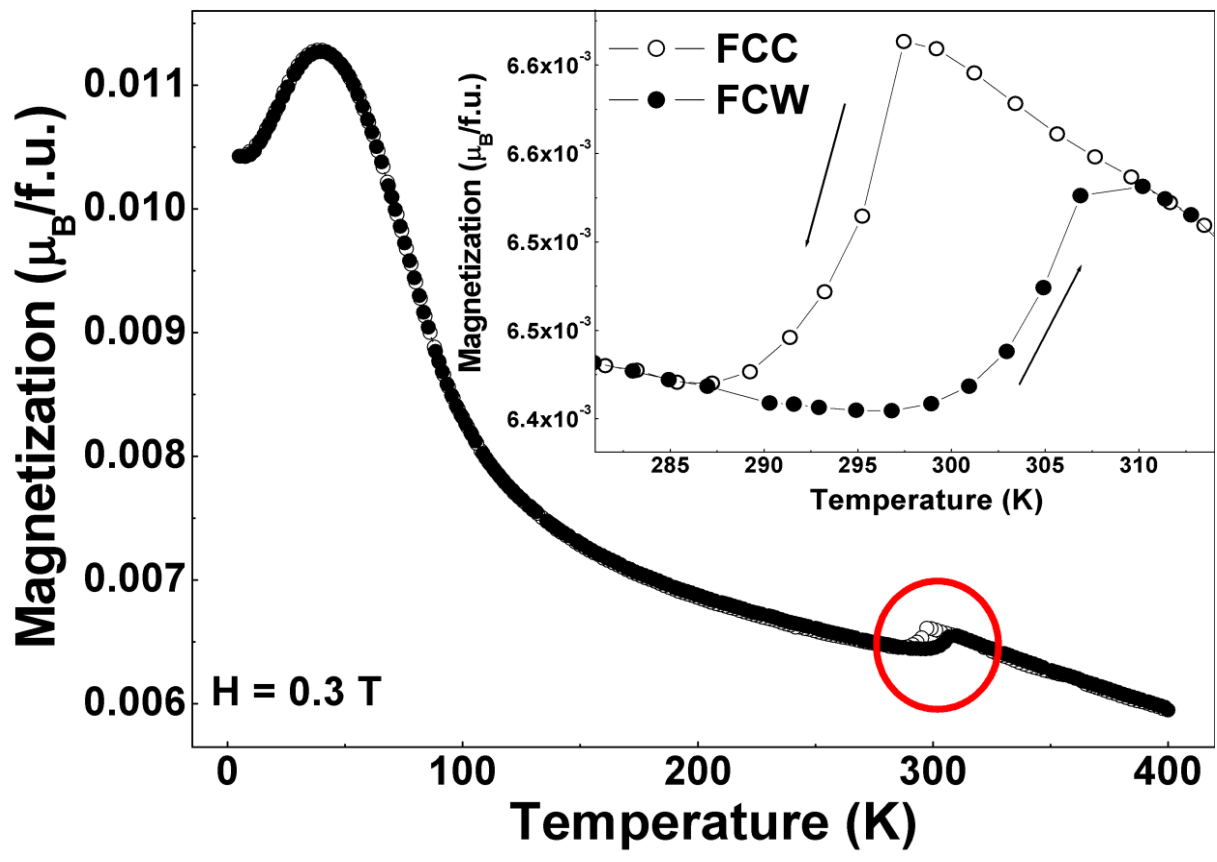

Figure 3

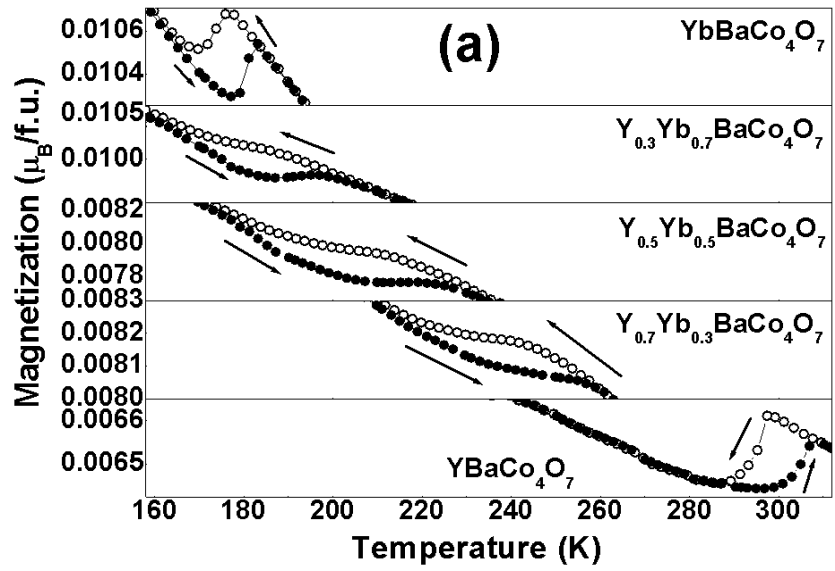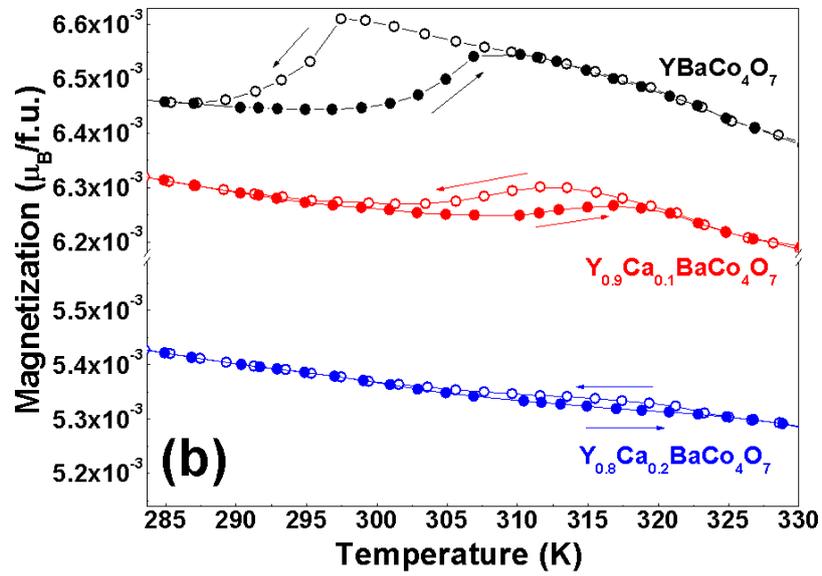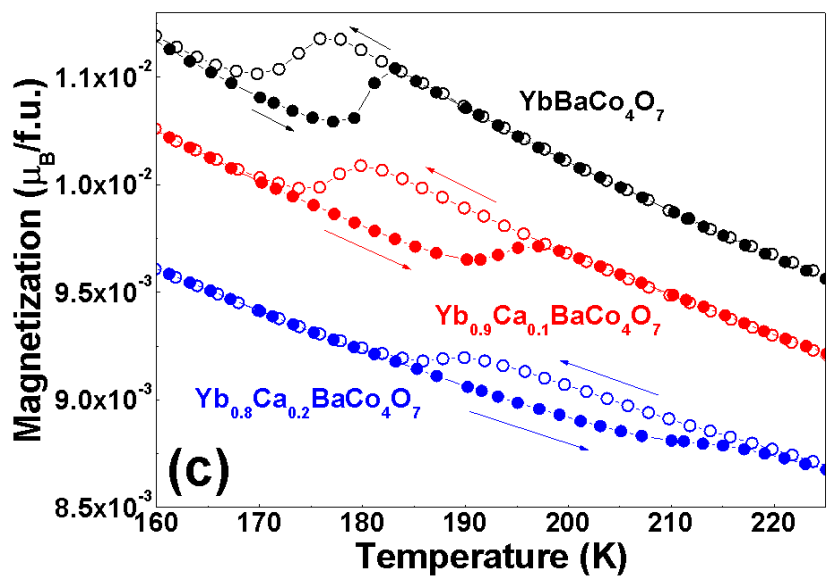

Figure 4

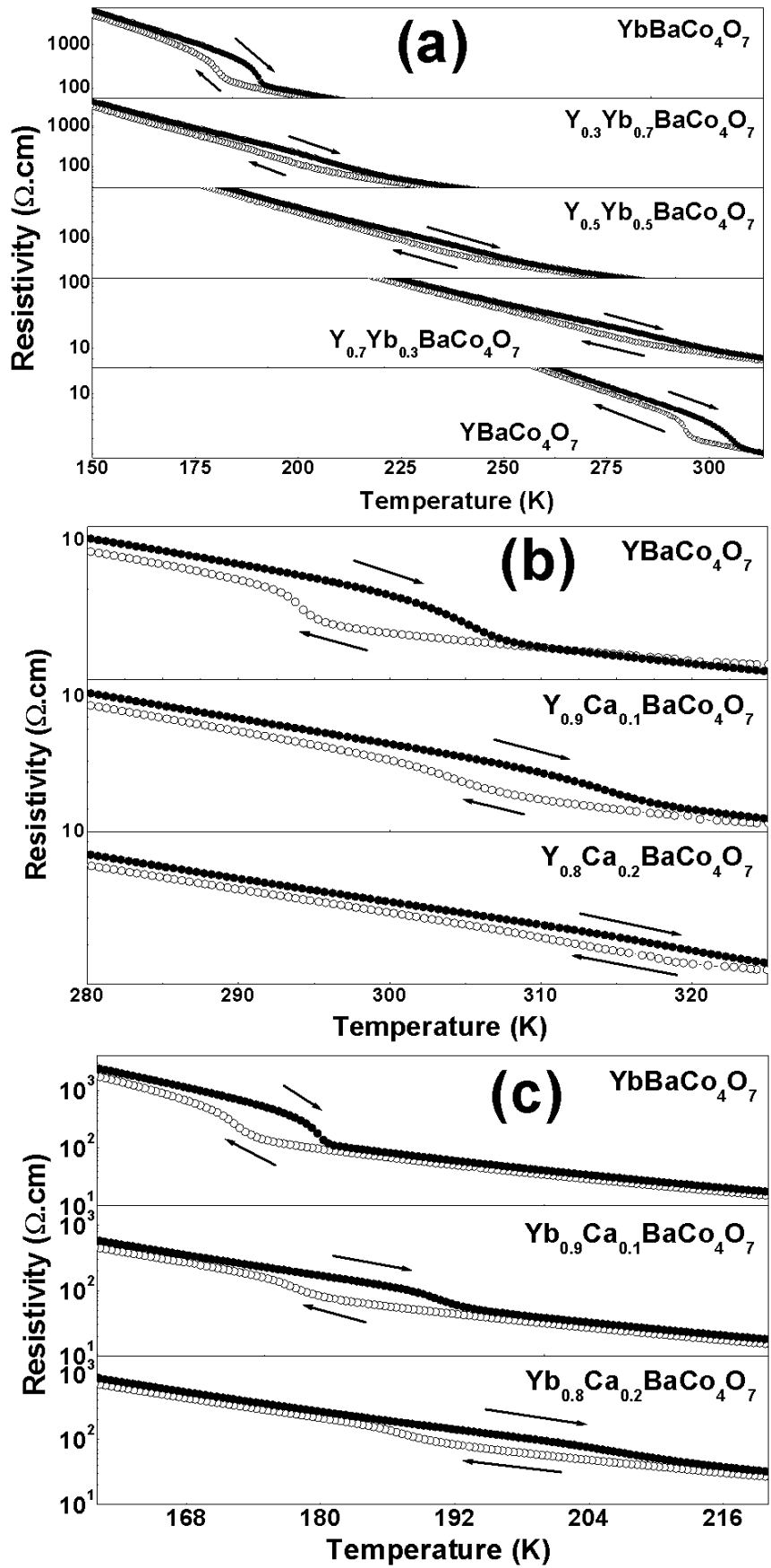

Figure 5

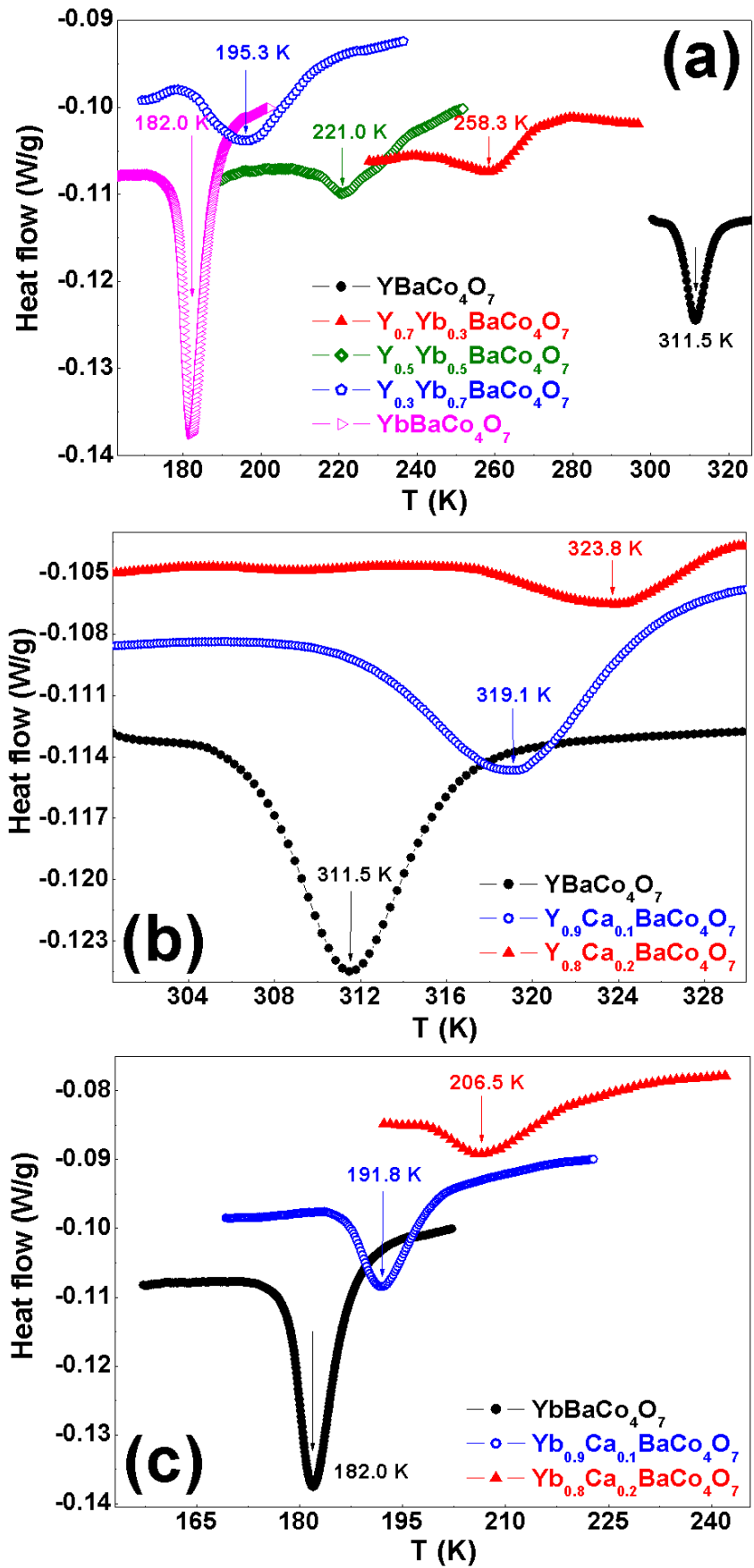

Figure 6

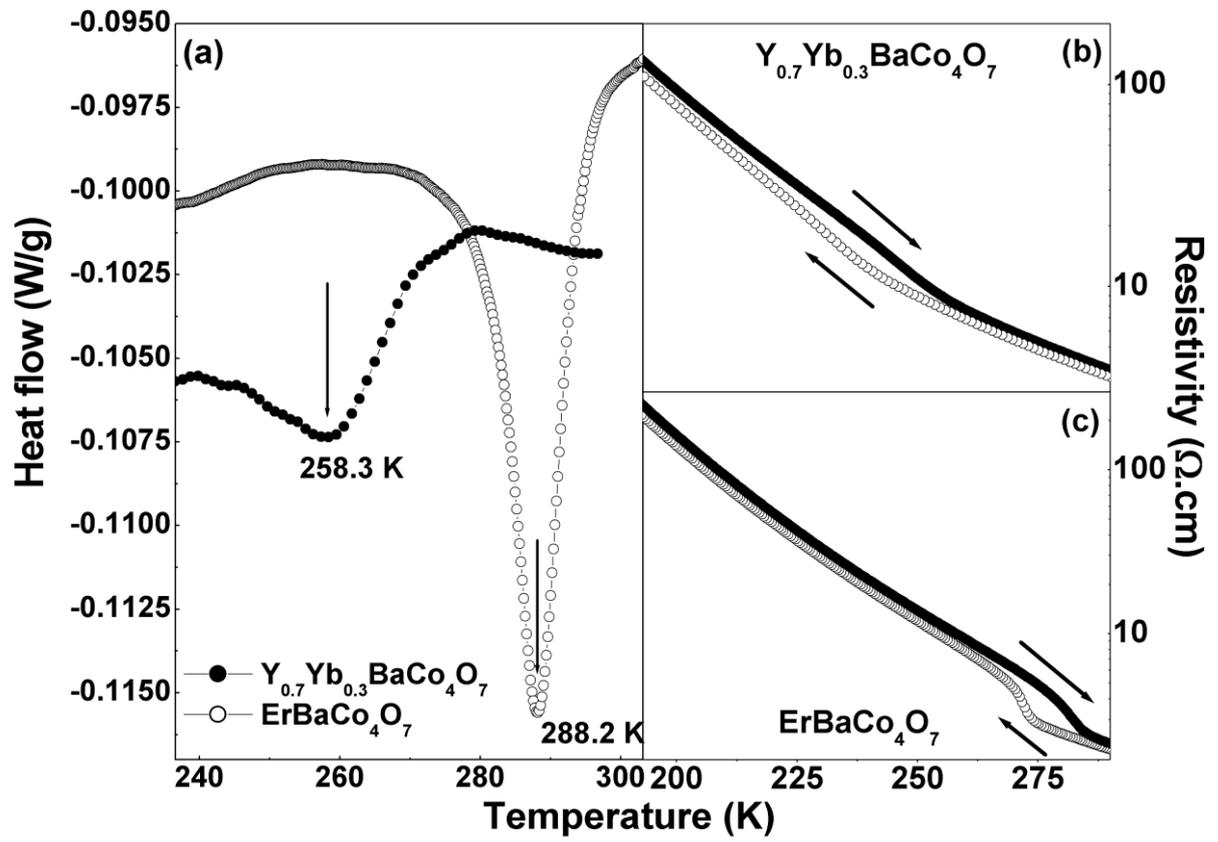

Figure 7